\documentclass[aps,pra,reprint,amsmath,amssymb,longbibliography,nofootinbib]{revtex4-2}

\usepackage{soul}
\usepackage[utf8]{inputenc}
\usepackage{bm,amsfonts,amsmath,amssymb,color}
\usepackage{dcolumn,xfrac}
\usepackage{graphicx}
\def\la{\;
\raise0.3ex\hbox{$<$\kern-0.75em\raise-1.1ex\hbox{$\sim$}}\; }
\def\ga{\;
\raise0.3ex\hbox{$>$\kern-0.75em\raise-1.1ex\hbox{$\sim$}}\; }

\begin{document}

\title{Acetaldehyde as a molecule for testing variations of
    electron-to-proton mass ratio}

% MDPI internal command: Title for citation in the left column

\author{J. S. Vorotyntseva$^{1,2}$}
\author{S. A. Levshakov$^{1}$} 
\author{M. G. Kozlov$^{2,3}$}

\affiliation{
$^{1}$ Ioffe Institute, Polytekhnicheskaya Str. 26, 194021 St.~Petersburg, Russia\\
$^{2}$ St.~Petersburg Electrotechnical University ``LETI'', Prof. Popov Str. 5, 197376 St.~Petersburg, Russia\\
$^{3}$ Petersburg Nuclear Physics Institute of NRC ``Kurchatov Institute'', Gatchina, Leningrad District, 188300, Russia
}

\begin{abstract}
We present the quantum-mechanical calculations of the
dimensionless  sensitivity coefficients $Q$ to small changes in the fundamental 
physical constant $\mu = m_e/m_p$~-- the electron-to-proton mass ratio~-- 
for a number of low-frequency (1--50 GHz) transitions of the acetaldehyde (CH$_3$CHO) molecule. 
The calculations show that  $Q$ 
varies in the range from 0.62 to 3.61. 
An example of the practical use of the CH$_3$CHO and CH$_3$OH lines tracing the same regions 
in three molecular clouds, located 
at large galactocentric distances ($D_{\scriptscriptstyle\rm GC} \sim 8$ kpc) is considered.
This results in a limit on the $\mu$ variations of
$\Delta\mu/\mu  = (0.1 \pm 0.4)\times 10^{-7}$
which is in line with previously obtained most stringent upper limits on changes in $\mu$
based on other molecules and methods. 
The limit obtained restricts hypothetical violations of the Einstein principle of the local
position invariance at the level of $4\times10^{-8}$ in the Galactic disk
at large galactocentric distances. 
\end{abstract}

\date{\today}

\maketitle

%-------------------------------Section 1
\section{Introduction}

\label{Sec1}
Complex organic molecules  are carbon-bearing molecules that consist of 
at least 6 atoms.
At astrophysical conditions these molecules are observed in the interstellar 
medium (ISM) and are present at all stages of star formation, from dense molecular 
clouds to protostellar, young stellar objects  and protoplanetary disks. 
This prevalence makes it possible to investigate the physical and chemical conditions 
throughout the star and planet formation processes using spectral
observations of such molecules. 

Among complex organic molecules, a special class can be distinguished~-- molecules with hindered internal motion. 
Currently, simpler 6-atom molecules (CH$_3$OH, CH$_3$SH) and more complex, up to 
12-atom molecule C$_2$H$_5$OCH$_3$, are detected in the ISM \cite{Kle19, Hollis, Req}. They are listed  in Table~\ref{T1}.

%-----------------------------Table 1
\begin{table}[h!]
%\centering
\caption{Complex organic molecules with hindered internal motion observed in the interstellar medium \cite{Kle19, Hollis, Req}.}
\label{T1}
\begin{tabular}{c c c c c}
\hline\\[-7pt]
Number of &\multicolumn{4}{c}{Molecular}\\
atoms&\multicolumn{4}{c}{formula}\\
\hline\\[-7pt]
6 & CH$_3$OH&CH$_3$SH&&\\
7 &CH$_3$CHO&CH$_3$NH$_2$&CH$_3$NCO&\\
8& HC(O)OCH$_3$&CH$_3$COOH&CH$_3$CHNH&\\
9& CH$_3$OCH$_3$&CH$_3$CONH$_2$&CH$_3$NHCHO&\\
10& CH$_3$COCH$_3$&CH$_3$CHCH$_2$O& CH$_3$OCH$_2$OH \\
 &  HOCH$_2$CH$_2$OH \\
11&CH$_3$C(O)OCH$_3$&&&\\
12&C$_2$H$_5$OCH$_3$&&&\\
\hline\\[-7pt]
\end{tabular}
\end{table}

Molecules with hindered  internal  motion have an important specificity: 
torsion-rotational transitions of such molecules, which fall within the microwave spectral range, 
have high sensitivity coefficients $Q$ to hypothetical variations in the electron-to-proton mass 
ratio $\mu=m_e/m_p$, individual for each transition.  This makes such molecules not only a probe of 
the physical and chemical conditions in star forming regions, but also a tool for testing fundamental physical laws,
such as, e.g., Einstein's principle of 
the local position invariance (LPI), which states that
non-gravitational measurements are independent of their location in space 
and time.  A violation of the LPI is predicted in a number of theories \cite{Uz25}. 

In general, sensitivity coefficients $Q$ have previously been calculated for many molecules. 
The first molecule for these purposes was molecular hydrogen H$_2$, for the Lyman and Werner  
electro-vibro-rotational
transitions of which the $Q$ values  turned out to be rather small, $Q \sim10^{-2}$ \cite{VL} as compared with pure rotational molecular transitions, having $Q = 1$
\cite{KL2013}.
It was also shown that the inversion transition ($J,K = 1,1$) of the ammonia molecule NH$_3$ 
has a rather high value of  $Q = +4.46$ \cite{FK}.

In 2012-2013, calculations were performed for methyl mercaptan molecule CH$_3$SH~-- for its various transitions, the values of $Q$ range from --12.2 to +14.8 \cite{J13}, and for methylamine CH$_3$NH$_2$ with --24 $\leq Q \leq$ +19 \cite{Il}.
For a number of complex organic molecules (CH$_3$CHO, CH$_3$CONH$_2$, CH$_3$OCOH, CH$_3$COOH), sensitivity coefficients for a few  transitions were calculated in \cite{J11} as well.

However, as it was shown by two independent methods in 2011, the highest values of the sensitivity coefficients have the transitions of methanol CH$_3$OH: $-53 \leq Q \leq +42$ \cite{LKR, J112}. 
Compared with the pioneering work on molecular hydrogen, the use of methanol transitions yields an efficiency 1000 times greater in estimates of $\mu$-variations.
Based on this fact, $Q$-coefficients have recently been  calculated for  methanol isotopologues, which showed that these molecules have even higher sensitivity coefficients than basic methanol: $-32 \leq Q \leq +78$ ($^{13}$CH$_3$OH),  $-109 \leq Q \leq +33$ (CH$_3$$^{18}$OH) , and   $-32 \leq Q \leq +25$, $-300 \leq Q \leq +73$,  $-44 \leq Q \leq +38$ for deuterated methanol CH$_3$OD, CD$_3$OH and CD$_3$OD, respectively \cite{VKL, VLK}.

The hypothetical variations in $\mu$ can be estimated from a pair of 
molecular transitions with different $Q$-values \cite{LKR}:
\begin{equation}
\frac{\Delta\mu}{\mu}= \frac{V_j-V_i}{c(Q_{i}-Q_{j})},
\label{Eq1}
\end{equation}
where $V_j$ and $V_i$ are the measured 
radial velocities in the local standard of rest (LSR) velocity system, $V_{\rm LSR}$,
of a pair of molecular transitions having sensitivity coefficients $Q_{j}$ and
$Q_{i}$, $c$ is the speed of light, and $\Delta\mu/\mu$  is the fractional difference between the
astronomical, $\mu_{\rm obs}$, and terrestrial, $\mu_{\rm lab}$, values of $\mu$: 
$\Delta\mu/\mu = (\mu_{\rm obs}-\mu_{\rm lab})/\mu_{\rm lab}$.
The radial velocity is defined according to the radio astronomical convention:
\begin{equation}
V_{\rm LSR} = c (1 - f_{\rm obs}/f_{\rm lab}),
\label{Eq1a}
\end{equation}
where $f_{\rm lab}$ and $f_{\rm obs}$ are the 
laboratory and observed transition frequencies, respectively. 

The tightest upper limits on $\Delta\mu/\mu$ in the Galactic disk at 
large galactocentric distances ($D_{\scriptscriptstyle\rm  GC} \ga 4$ kpc) 
are obtained with observations of the NH$_3$\, (1,1) inversion transition 
in combination with pure rotational transitions  of HC$_3$N, HC$_5$N, and HC$_7$N: 
$\Delta\mu/\mu < 7 \times 10^{-9}$ \cite{L2013}; and with methanol lines (CH$_3$OH 
and $^{13}$CH$_3$OH): $\Delta\mu/\mu < (1 - 3) \times 10^{-8}$ \cite{VLK, Dap, L22, Ell, VL2024}\footnote{All 
upper limits on $\Delta\mu/\mu$ throughout the paper are given at the $1\sigma$ confidence level.}.

It should be noted that in the vicinity of the Galactic center
$(D_{\scriptscriptstyle\rm GC} \simeq 0.1$ kpc), 
which contains a super massive black hole 
$(M_{\scriptscriptstyle\rm BH} \sim 4\times10^6M_\odot)$
and a
massive complex of molecular clouds Sagittarius (Sgr)\,B2
$(M_{\scriptscriptstyle\rm Sgr} \sim 6\times10^7M_\odot)$,
an indication of a reduced value of $\mu$ 
has recently been found from the microwave spectra of methanol
(CH$_3$OH) observed in the hot core Sgr\,B2(N):
$\Delta\mu/\mu = (-4.2\pm0.7)\times10^{-7}$  \cite{VL2025}~-- based on
the {\it Herschel} space telescope observations\footnote{{\it Herschel} is an ESA
space observatory with science instruments provided by European-led Principal
Investigator consortia and with important participation from NASA. }, 
and
$\Delta\mu/\mu = (-2.1\pm0.6)\times10^{-7}$  \cite{VLH2025}~-- based on 
the IRAM 30-m telescope observations\footnote{The Institute for Radio Astronomy 
in the Millimeter Range (IRAM) is an international research institute and Europe's leading
center for radio astronomy. }.  
At the same time, the identical with Sgr\,B2(N) set of methanol lines observed 
with {\it Herschel} toward another complex of molecular clouds Orion-KL 
$(D_{\scriptscriptstyle\rm GC} \simeq 8$ kpc) yields only  an upper limit on
$\Delta\mu/\mu = (-0.5\pm0.6)\times10^{-7}$  \cite{VL2025}.

This shows that
the search for new molecules that are ubiquitous in space  
and that can be used to assess $\mu$
remains an urgent task  to confirm previously obtained evaluations of $\Delta\mu/\mu$ and to mitigate possible systematic effects.

One of such molecules is acetaldehyde, CH$_3$CHO,
which is widespread in the different environments of the ISM.
The sensitivity coefficients for its  selected eight microwave transitions  
from the frequency range $\Delta f = 1.8 - 47.8$ GHz were
calculated in \cite{J11}.
 
In present paper, we extended the list of acetaldehyde lines from the low frequency range 
$\Delta f = 1 - 50$ GHz and calculated the corresponding $Q$-values using a different method  based on a simpler structure of the Hamiltonian. 
The results obtained are given in Section~\ref{Sec2}.
In Section~\ref{Sec3},   we discuss an example of the $\Delta\mu/\mu$  estimation from 
a combined sample of methanol and acetaldehyde lines detected in spectra of three molecular clouds.
And finally, in Section~\ref{Sec4}, we conclude.

%-----------------------------Section 2
\section{Sensitivity coefficients of selected transition in acetaldehyde (CH$_3$CHO) }
\label{Sec2}

\subsection{General remarks}
\label{SSec2-1}

Acetaldehyde (CH$_3$CHO) is a complex organic molecule similar in structure to methanol CH$_3$OH, which is also characterized by hindered internal rotation.
CH$_3$CHO consists of 7 atoms -- three hydrogens H and carbon C in the methyl group CH$_3$, carbon C, oxygen O and hydrogen H in the CHO group. The internal rotation here is related to the quantum mechanical tunneling effect. The methyl group CH$_3$ can undergo torsional vibrations relative to the CHO group, where the hydrogen atom can be  in one of three possible positions with equal energies.

There are two types of acetaldehyde that differ in symmetry properties: $A$-type, for which the total nuclear spin of the methyl group is 
$I_{{\scriptscriptstyle\rm CH}_3}$ = 3/2, and $E$-type, with 
$I_{{\scriptscriptstyle\rm CH}_3}$ = 1/2. 

It is worth noting that transitions between $A$- and $E$-acetaldehyde are not observed, since radiative and collisional transitions
between the two types with proton spin flips are unlikely due to the extremely
low efficiency of interaction of nuclear spins with magnetic and electric fields.

\begin{figure}[h!]
%\vspace{-2cm}
\centering
\includegraphics[width=0.5\textwidth]{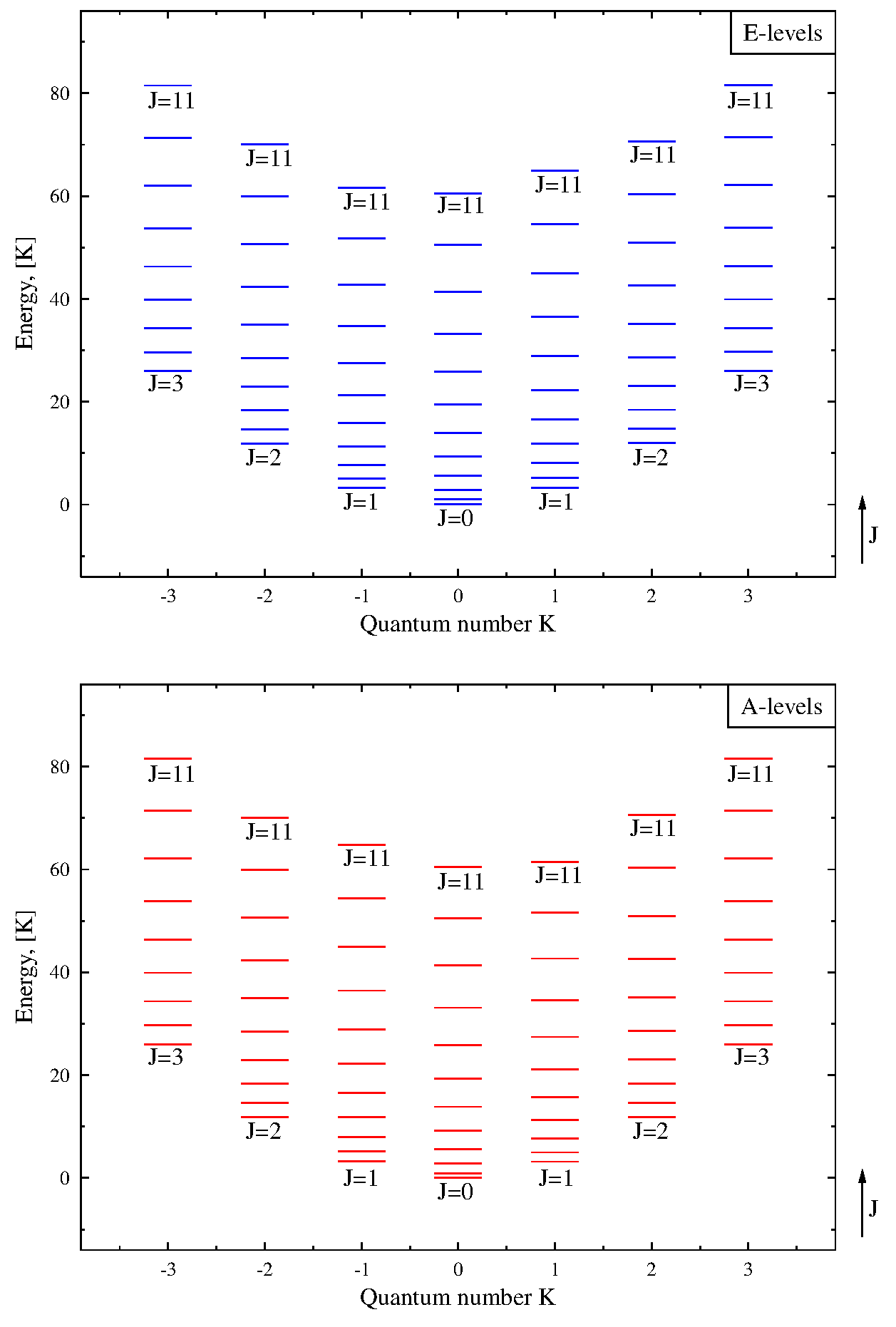}
%\includegraphics[width=0.5\textwidth]{fig-5}
%\vspace{-2.0cm}
\caption{\small Partial rotational level diagrams of $A$- and $E$- levels of acetaldehyde CH$_3$CHO
in the torsional ground state ($v_t=0$). 
%Numbers
% below the horizontal bars are the values of the rotational angular
% momenta $J$.
The energy of the ground torsion state of the 
$E$-type acetaldehyde is 0.097~K.
 }
\label{F1}
\end{figure}

Figure~\ref{F1} shows a diagram of the $A$- and $E$- levels of the CH$_3$CHO molecule.
Note that the energy of the ground torsion state of the 
$E$-type acetaldehyde ($v_t = 0, J = 0$) differs from zero and is 0.097~K.
A similar characteristic is also observed for the $E$-type methanol.

%-----------------------------------Figure 1

\subsection{Calculating procedure}
\label{SSec2-2}

To calculate the sensitivity coefficients $Q$, the procedure described in \cite{LKR} is used. This procedure is based on a simple and physically transparent model of the effective Hamiltonian \cite{RF}, containing only 7 spectroscopic parameters which have a clear physical insight: three rotation
 parameters $A,B,C$, one parameter $D$, describing interaction
 of the internal rotation with overall rotation, the kinetic coefficient $F$, the depth of the threefold symmetric torsion potential $V_3$, and a dimensionless parameter $\rho$~-- the internal rotation interaction constant. The listed parameters can be calculated using the following formulas:
\begin{equation}
A = \frac{1}{2}\hbar^2 \left(\frac{I_a+I_b}{I_aI_b-I^2_{ab}}-\frac{I_b}{I^2_b+I^2_{ab}}\right),
\label{Eq2}
\end{equation}

\begin{equation}
B = \frac{1}{2}\hbar^2 \frac{I_b}{I^2_b+I^2_{ab}},
\label{Eq3}
\end{equation}

\begin{equation}
C = \frac{1}{2}\hbar^2 \frac{1}{I_c},
\label{Eq4}
\end{equation}

\begin{equation}
D = \frac{1}{2}\hbar^2 \frac{I_{ab}}{I^2_b+I^2_{ab}},
\label{Eq5}
\end{equation}

\begin{equation}
F = \frac{1}{2}\hbar^2 \frac{I_a I_b - I^2_{ab}}{I_{a2}(I_{a1}I_{b}-I^2_{ab})},
\label{Eq6}
\end{equation}

\begin{equation}
\rho =  \frac{I_{a2}\sqrt{I^2_b+I^2_{ab}}}{I_{a}I_{b}-I^2_{ab}}.
\label{Eq7}
\end{equation}

Here $\hbar=h/2\pi$; $I_a$, $I_b$, $I_c$ are the moments of inertia. $I_{ab}$
is the product of inertia about the $a$- and $b$-axes in the $a,b,c$-axis system whose $a$-axis is parallel
to the internal rotation axis (assumed to be that of the methyl top), the $c$-axis perpendicular to the CHO plane.
$I_{a2}$ is the axial moment of inertia of the methyl group, and $I_{a1}$ is that of the framework 
(the CHO group), the sum of which determines $I_{a}$:
$I_{a} = I_{a1} + I_{a2}$.

The moments of inertia, displayed in Table \ref{T2}, are assumed to be known from \cite{NB}.

%-----------------------------Table 2
\begin{table}[h!]
\centering
\caption{Moments of inertia (in units amu$\cdot$\AA$^2$) for the acetaldehyde CH$_3$CHO \cite{NB}.}
\label{T2}
\begin{tabular}{c c c c}
\hline\\[-8pt]
$I_{a}$ & $I_{b}$  & $I_{ab}^\ast$&$I_{c}$\\
\hline\\[-2pt]
15.6230 &  43.0165&--15.1467&55.6095\\
\hline\\[-8pt]
\multicolumn{4}{l}{\footnotesize  ${^\ast}I_{ab} = I_{ba}$ in \cite{NB}.}
\end{tabular}
\end{table}

Table~\ref{T3} shows the values of the spectroscopic constants calculated from the moments of inertia in accordance with the formulas (\ref{Eq2})-(\ref{Eq5}). Table \ref{T3} also contains an asymmetry parameter $\kappa$ which is defined as \cite{TowSh}:
\begin{equation}
\kappa =  \frac{2B-A-C}{A-C}\ ,
\label{Eq8}
\end{equation}
 $\kappa$ is equal to 1 or --1 when the molecule is a symmetric top.
In Table \ref{T3}, the methanol parameters from \cite{VLK} are 
also listed for comparison.

%-----------------------------Table 3
\begin{table}[h!tb]
\centering
\caption{Spectroscopic parameters of the effective Hamiltonian 
and the asymmetry parameter $\kappa$ 
for CH$_3$CHO and CH$_3$OH -- for comparison.}
\label{T3}
\begin{tabular}{ l r@.l r@.l}
\hline\\[-5pt]
 & \multicolumn{2}{c}{CH$_3$CHO} & \multicolumn{2}{c}{CH$_3$OH}\\
\hline\\[-7pt]
$A$ (cm$^{-1}$) & 1&8846 & 4&2556 \\
$B$ (cm$^{-1}$) & 0&3487  & 0&8232   \\
$C$ (cm$^{-1}$) & 0&3031  & 0&7928 \\
$D$ (cm$^{-1}$) & $-$0&1228   &$-0$&0026  \\
$F$ (cm$^{-1}$)$^{\ast}$ & 7&6559 & 27&7518\\
$V_3$ (cm$^{-1}$)$^{\ast}$ & 407&9& 375&6 \\
$\rho^{\ast}$ & 0&3291& 0&8109 \\
$\kappa$ & $-0$&942 & $-0$&982 \\
\hline\\[-8pt]
\multicolumn{5}{l}{\footnotesize $^\ast$Values for acetaldehyde CH$_3$CHO }\\
\multicolumn{5}{l}{\footnotesize are taken from \cite{Kle96}.}
\end{tabular}
\end{table}

The calculated spectroscopic parameters were used to determine the Hamiltonian matrix,
the non-vanishing elements of which are listed in \cite{VKL}
for specific values of the rotational angular momentum $J$.
Diagonalization of this matrix gives simultaneously
the eigenvalues (energies) and eigenfunctions for the $A$- and $E$-type acetaldehyde. 
The dependence of the found eigenvalues $E_i$ on $\Delta\mu/\mu$
defines the sensitivity coefficient $Q$ through the following relations \cite{LKR}:
\begin{equation}
\Delta E_i = q_i\frac{\Delta\mu}{\mu},
\label{Eq9}
\end{equation}
and
\begin{equation}
Q = \frac{q_{\scriptscriptstyle u} - q_{\scriptscriptstyle \ell}}{hf}.
\label{Eq10}
\end{equation}
Here $\Delta E_i$ is the energy shift due to non-zero $\Delta\mu/\mu$,
$q_i$ is the so-called $q$-factor, individual for each level $E_i$,
which shows a response of this level  to a small changes in $\mu$, when
$|\Delta\mu/\mu| \ll 1$,  $f$ is the laboratory frequency of a given transition,
and $h$ is Planck's constant.
The subscripts $u$ and $\ell$ denote the upper and lower levels, respectively.
This functional dependence is assessed through the diagonalization of the 
Hamiltonian matrix for three sets of parameters that correspond to
$\mu = \mu_{\scriptscriptstyle 0}$ and 
$\mu = \mu_{\scriptscriptstyle 0}(1 \pm \varepsilon)$, where $\varepsilon$
is equal to 0.001 or 0.0001, and $\mu_{\scriptscriptstyle 0} \equiv 
\mu_{\scriptscriptstyle\rm lab}$ \cite{LKR}.

\subsection{Calculated sensitivity coefficients}
\label{SSec2-3}

For calculations we selected molecular transitions with $0 \le J \le 11$
and $|\Delta K| = 1$,
since transitions with $\Delta K = 0$ are pure rotational
and have $Q \approx 1$.
However, for a number of lines with $\Delta K = 0, 1, 2$, observed in molecular clouds, we also calculated the sensitivity coefficients and their errors.
Our calculations 
cover the low frequency range $\Delta f = 1-50$~GHz which contains 
CH$_3$CHO transitions most sensitive to $\mu$-variations.

The results of our calculations are reported in
Tables~\ref{T4}-\ref{T6}. In the first column of Table~\ref{T4}, the quantum numbers of the molecular transitions are shown~-- the total angular momentum $J$ and its projection $K$ onto the molecular axis
for the upper $u$ and lower $\ell$ states, 
the second column lists the laboratory frequencies $f$ from \cite{Kle96}, the third column displays the calculated sensitivity coefficients $Q$
and their errors, and in the last column the line strengths (in D$^2$) are shown.
The structure of Table~\ref{T5} is similar, but instead of the measured  laboratory frequency, the predicted frequency is shown. Table~\ref{T6} contains the same data as Table~\ref{T4}, but only for transitions observed in the ISM 
with the references and object names listed in the last column.

The errors of the $Q$ values shown in parentheses are calculated using the same procedure as for methanol \cite{VKL, VLK}.
From Eqs. (\ref{Eq2})-(\ref{Eq6}) it can be seen that the Hamiltonian parameters $A - F$ are inversely proportional to the moments of inertia, which provides a linear dependence of $\mu$ (each parameter is scaled as $\mu^{-1}$). At the same time, potential barrier $V_3$ is independent of $\mu$ and is scaled as $\mu^0$. However, because of the weak dependence of the internuclear distances on $\mu$, due to the vibrational and centrifugal distortions, the deviation from the above mentioned scalings can be about 2\% \cite{LKR}. The resulting errors of $Q$ are determined by the quadratic sum of the errors caused by changes in the values of each parameter within the 2\% uncertainty.

It is worth noting that
Tables~\ref{T4}-\ref{T6} show CH$_3$CHO transitions whose $Q$ coefficients have relative errors of less than 50\%.

It is also necessary to pay attention to the fact that the error calculation procedure described in \cite{VKL, VLK} yields a majorizing estimate of uncertainty in $Q$, since it includes the contribution from the change in each of the rotational constants.
At the same time, it is not necessary that all constants will change simultaneously, therefore the errors of $Q$ should be taken as the maximum possible, although in reality they may have a smaller value.
 
For some lines listed in Table~\ref{T6}, sensitivity coefficients and their errors were calculated by another method in \cite{J11}.
Both calculations show good agreement within the $\pm1\sigma$ confidence interval
(see footnote to Table~\ref{T6}).

%-----------------------------Table 4
\begin{table}[h!]
%\centering
\caption{Calculated sensitivity coefficients $Q$ for 
the low-frequency ($\Delta f = 1-50$ GHz)
torsion-rotation transitions ($\Delta K=\pm 1$) in CH$_3$CHO
measured in laboratory \cite{Kle96}.
Given in parenthesis are errors  in the last digits.}
\label{T4}
\begin{tabular}{l r@.l r@.l c}
\hline
\\[-8pt]
\multicolumn{1}{c}{$J_{K_u} \to J_{K_{\ell}}$} & 
\multicolumn{2}{c}{$f$, MHz} & 
\multicolumn{2}{c}{$Q$} & 
$S\mu^2_{\scriptscriptstyle D}$\\
\hline\\[-7pt]
{$1_{-1} \to 2_{0}E$}& 7391&311& 0&62(18)&0.4618\\  
{$1_{1} \to 2_{0}E$}& 9240&920& 1&22(14)&0.1265\\  
{$7_{2} \to 8_{1}A^+$}& 9542&776& 1&4(8)&1.1288\\  
{$3_{0} \to 2_{1}A^+$}& 12014&990& 0&88(13)&1.2070\\ 
{$3_{0} \to 2_{-1}E$}& 12635&239& 1&13(12)&1.1435\\ 
{$10_{3} \to 11_{2}E$}& 14456&880& 1&2(6)&0.9221\\ 
{$10_{3} \to 11_{2}A^+$}& 14691&111& 1&3(6)&1.8549\\  
{$6_{2} \to 7_{1}A^+$}& 23364&771& 1&2(3)&0.9936\\ 
{$10_{-3} \to 11_{-2}E$}& 24859&393& 0&9(3)&0.8897\\  
{$10_{3} \to 11_{2}A^-$}& 27173&880& 1&2(2)&1.7823\\ 
{$8_{1} \to 7_{2}E$}& 29310&400& 0&92(19)&0.2753\\   
{$8_{1} \to 7_{2}A^-$}& 30941&333& 0&9(2)&1.7519\\ 
{$4_{0} \to 3_{1}A^+$}& 32709&185& 0&96(6)&1.8589\\ 
{$4_{0} \to 3_{-1}E$}& 33236&469& 1&03(6)&1.8268\\  
{$5_{-2} \to 6_{-1}E$}& 35472&152&0&85(14)&0.3298\\ 
{$9_{3} \to 10_{2}A^+$}& 36613&186& 1&1(2)&1.5908\\ 
{$4_{2} \to 5_{1}A^-$}& 37516&317& 1&10(12)&0.7415\\ 
{$5_{2} \to 6_{1}A^+$}& 38138&436& 1&09(13)&0.8226\\ 
{$2_{-1} \to 2_{0}E$}& 45078&240& 0&96(3)&0.1320\\ 
{$9_{3} \to 10_{2}A^-$}& 45344&590& 1&10(14)&1.5489\\ 
{$1_{-1} \to 1_{0}E$}& 45897&330& 0&94(2)&0.3593\\ 
\hline\\[-5pt]
\end{tabular}
\end{table}

%-----------------------------Table 5
\begin{table}[h!]
%\centering
\caption{Calculated sensitivity coefficients $Q$ for 
the predicted in \cite{Kle96}
low-frequency ($\Delta f = 1-50$ GHz)
torsion-rotation transitions ($\Delta K=\pm 1$) in CH$_3$CHO.
Given in parenthesis are errors  in the last digits.}
\label{T5}
\begin{tabular}{l r@.l r@.l c}
\hline
\\[-8pt]
\multicolumn{1}{c}{$J_{K_u} \to J_{K_{\ell}}$} & 
\multicolumn{2}{c}{$f$, MHz} & 
\multicolumn{2}{c}{$Q$} & 
$S\mu^2_{\scriptscriptstyle D}$\\
\hline\\[-7pt]
{$9_{-1} \to 8_{-2}E$}& 8397&01& 1&4(9)&0.0959\\
{$6_{-2} \to 7_{-1}E$}& 20274&43& 0&8(3)&0.2790\\
{$11_{-1} \to 11_{0}E$}& 21203&544& 1&0(2)&0.0005\\
{$10_{-1} \to 9_{-2}E$}& 21914&175& 1&12(4)&0.0465\\
{$10_{-1} \to 10_{0}E$}& 24270&204& 0&99(18)&0.0008\\
{$4_{0} \to 3_{1}E$}& 26691&022& 0&99(7)&0.0255\\
{$9_{-1} \to 9_{0}E$}& 27432&732& 0&99(15)&0.0013\\
{$8_{-1} \to 8_{0}E$}& 30609&589& 0&99(12)&0.0021\\
{$7_{-1} \to 7_{0}E$}& 33712&067& 0&99(9)&0.0035\\
{$11_{-1} \to 10_{-2}E$}& 34868&791& 1&1(3)&0.0224\\
{$6_{-1} \to 6_{0}E$}& 36649&808& 0&99(7)&0.0060\\
{$4_{2} \to 5_{1}E$}& 37641&748& 1&09(11)&0.3589\\
{$5_{-1} \to 5_{0}E$}& 39335&472& 0&99(5)&0.0110\\
{$4_{-1} \to 4_{0}E$}& 41687&673& 0&98(4)&0.0219\\
{$9_{-3} \to 10_{-2}E$}& 43113&861& 0&92(15)&0.8907\\
{$5_{0} \to 4_{1}E$}& 43539&148& 1&00(5)&0.0128\\
{$3_{-1} \to 3_{0}E$}& 43629&973& 0&98(3)&0.0492\\
{$2_{1} \to 2_{0}E$}& 48603&477& 1&02(2)&2.6797\\
{$2_{1} \to 2_{0}A^{-+}$}& 48902&831& 1&03(2)&2.8232\\
\hline\\[-5pt]
\end{tabular}
\end{table}

%-----------------------------Table 6
\begin{table}[h!]
%\centering
\caption{Calculated sensitivity coefficients $Q$ for 
the low-frequency ($\Delta f = 1-50$ GHz)
torsion-rotation transitions $(\Delta K = 0,1,2)$ 
in CH$_3$CHO observed in the ISM.
Given in parenthesis are errors in the last digits.}
\label{T6}
\begin{tabular}{l r@.l r@.l c r}
\hline\\[-7pt]
\multicolumn{1}{c}{$J_{K_u} \to J_{K_{\ell}}$} & 
\multicolumn{2}{c}{$f$, MHz} & 
\multicolumn{2}{c}{$Q$}  & $S\mu^2_{\scriptscriptstyle D}$&Object and ref.\\
\hline\\[-7pt]
{$1_{1} \to 1_{1}A^{-+}$}& 1065&075& 1&0(3)&9.4857&SgrB2$^\dagger$ \cite{Bell}\\ 
{$1_{1} \to 1_{-1}E$}& 1849&634& 3&61(12)$^a$&3.1329&SgrB2(N) \cite{Hol}\\
{$2_{1} \to 2_{1}A^{-+}$}& 3195&167& 1&0(3)&5.2690&SgrB2 \cite{Bell,Fou} \\
{$3_{1} \to 3_{1}A^{-+}$}& 6390&085& 1&0(3)&3.6878&SgrB2 \cite{Bell}\\ 
{$1_{1} \to 2_{0}A^+$}& 8243&476& 1&17(16)$^b$&0.5906&SgrB2(N) \cite{Hol}\\
{$4_{1} \to 4_{1}A^{-+}$}& 10648&419& 1&0(3)&2.8450&SgrB2 \cite{Bell}\\ 
{$1_{0} \to 0_{0}E$}& 19262&140& 1&000(14)$^c$&6.3282& SgrB2, 
TMC-1$^\diamond$,\\
&  \multicolumn{2}{c}{ } &  \multicolumn{2}{c}{ } & & 
L134N$^\times$  \cite{Matt85}\\ 
{$1_{0} \to 0_{0}A^{+}$}& 19265&133& 1&000(14)$^d$&6.3239&SgrB2, TMC-1, \\
& \multicolumn{2}{c}{ } &  \multicolumn{2}{c}{ }&& L134N  \cite{Matt85}\\ 
{$2_{1} \to 1_{1}A^+$}& 37464&168& 1&00(2)&9.4864&PMCs$^{\square}$ \cite{Sci24}\\ 
{$2_{-1} \to 1_{-1}E$}& 37686&868& 1&00(8)&8.8712&PMCs \cite{Sci24}\\ 
{$2_{0} \to 1_{0}E$}& 38505&999& 1&000(14)$^e$&12.6551&TMC-1  \cite{Kaif, Soma}\\ 
{$2_{0} \to 1_{0}A^{+}$}& 38512&113& 1&000(14)$^f$&12.6465&TMC-1 \cite{Kaif, Soma}\\ 
{$2_{1} \to 1_{1}E$}& 39362&504& 1&00(18)&8.8688&PMCs \cite{Sci24}\\ 
{$2_{1} \to 1_{1}A^-$}& 39594&287& 1&00(2)&9.4865&PMCs \cite{Sci24}\\ 
{$1_{1} \to 1_{0}E$}& 47746&980& 1&04(2)$^g$&1.3472&SgrB2(M) \cite{Sai} \\
{$1_{1} \to 1_{0}A^{-+}$}& 47820&668$^{\ast}$& 1&03(2)$^h$&1.7133&
SgrB2(M) \cite{Sai}\\ 
\hline\\[-5pt]
\multicolumn{7}{l}{\footnotesize $^\ast$The line was not observed in laboratory.
$^\dagger$Sagittarius molecular }\\
\multicolumn{7}{l}{\footnotesize clouds, Sgr;  $^\diamond$Taurus molecular clouds, TMC; $^\times$L refers to the } \\
\multicolumn{7}{l}{\footnotesize Lynds' catalog of dark nebulae;
$^{\square}$Perseus molecular clouds, PMC.}\\
\multicolumn{7}{l}{\footnotesize   $Q$ from \cite{J11}: $^a$3.7(2), 
$^b$1.11(6), $^c$1.00(5), $^d$1.00(5),  $^e$1.00(5), $^f$1.00(5),}\\
\multicolumn{7}{l}{\footnotesize  $^g$1.03(5),  $^h$1.02(5). }
\end{tabular}
\end{table}

Analyzing the data obtained from Tables~\ref{T4}-\ref{T6}, it can be said that the CH$_3$CHO molecule does not exhibit very  high sensitivity to possible changes
 in $\mu$: most transitions have $Q\sim 1$ and all of them have positive $Q$.
The explanation for this is as follows. 
In methanol, for example,
internal rotation is associated with the movement of only hydrogen, but here both hydrogen and oxygen rotate.
Such movement is much more hindered, and therefore the spectrum
of acetaldehyde  is much closer to purely rotational. 
Accordingly, the sensitivity coefficients $Q$ are closer to unity. In the Hamiltonian, this is manifested as follows: in methanol, the kinetic  coefficient $F$ is much higher than that in acetaldehyde ($\approx$ 7.6 in acetaldehyde vs $\approx$ 28 in methanol).
Moreover, although acetaldehyde is a fairly common molecule and is observed in many astronomical objects, the stronger lines observed are purely rotational and are essentially the only ones observed in space; the transitions with $\Delta K \not= 0$ in 
Tables~\ref{T4}-\ref{T6} are quite weak.
Therefore, it is difficult to estimate the variations in $\mu$ using this molecule alone. However, this is possible in combination with methanol lines, as in the
mentioned above ammonia method \cite{FK, L2013}.
But this approach would require careful analysis of the spectral line shapes or other data confirming the origin of the transitions of both molecules in the same region.
That such spatial coexistence between CH$_3$CHO and CH$_3$OH 
does take place is confirmed by direct observations of similar line profiles of these molecules in, e.g., Taurus Molecular Cloud-1 \cite{Soma}.

In conclusion of this section, we 
note that in addition to the above-mentioned transitions, which fall within the 
frequency range $\Delta f = 1-50$ GHz  from Table~\ref{T6}, purely rotational lines of acetaldehyde were also observed at higher frequencies.
Lines from the range $\Delta f = 70-120$ GHz were detected toward 
a good deal of molecular clouds in the Galactic disk 
\cite{Sci24, Bac, Bus, Sci20, Cod, Cer, Bhat, Char, Sci21, Nagy, Vas, Berg, Obe}.
Lines at even  higher frequencies (200--350 GHz) were detected in
\cite{Bhat, Hsi, Fay, Lyk}.
Also pure rotational transitions of CH$_3$CHO  (50--80 GHz) were observed  at the
cosmological  redshift $z = 0.89$ \cite{Mull11}.
All such lines have sensitivity coefficients close to unity, $Q \approx 1$.

%---------------------------------------------------------------Section 3

\section{Estimating $\Delta\mu/\mu$ using CH$_3$CHO and CH$_3$OH}
\label{Sec3}

In this section, we give an example of $\Delta\mu/\mu$ 
 estimation from a comparison of torsion-rotation transitions in CH$_3$OH and purely rotational transitions in CH$_3$CHO, which have different sensitivity coefficients $Q$.
Of course, among the methanol lines themselves there are purely rotational transitions with $Q=1$ and such estimates could be made based on the lines of one molecule.
However, for precision $\Delta\mu/\mu$ measurements, it is desirable to use close lines from narrow spectral ranges in which groups of methanol lines with significantly different $Q$ are absent.
In such cases, the acetaldehyde lines with $Q \approx 1$ could be used as references.

If the lines of methanol and acetaldehyde
trace the same spatial region, they should have approximately the same excitation temperature and approximately equal widths ($FWHM$).
The corresponding spectra are published in \cite{Bhat, data}, where the results of observations of many organic molecules, including CH$_3$CHO and CH$_3$OH, are presented in the direction of the three molecular clouds L1544, 
Barnard-1, and IRAS4A\footnote{L1544 is a protostellar system from the Lynds'
catalogue of dark nebular,  located in the Taurus molecular cloud complex. 
Barnard-1 is a dark nebula and IRAS4A is a protostellar system both embedded in the 
Perseus molecular cloud complex.  IRAS stands for Infrared Astronomical Satellite.}. 
The observations were carried out on the IRAM 30-m  telescope with a spectral resolution of 200 kHz in several spectral bands: 3 mm (80--116 GHz), 2 mm (130--170 GHz), and 1.3 mm (200--276 GHz).
  
For the estimates of $\Delta\mu/\mu$,
we selected methanol and acetaldehyde lines, satisfying the following criteria: ($i$) the lines should fall into the same spectral range of observations to exclude possible systematic shifts between different settings; ($ii$) the transitions should have approximately the same upper level energies (the difference is no more than 30 K); ($iii$)  the methanol lines should have $|Q| >1$, since the strongest acetaldehyde lines are purely rotational; and ($iv$) the laboratory frequencies should be
known from the measurements.

The molecular clouds with emission lines that meet the listed criteria are shown in
Table~\ref{T7}, where the first column indicates the object name, the second column shows the molecule, the third and fourth columns display the quantum numbers $J$ and $K$ for the upper $u$ and 
lower $\ell$ levels, the laboratory frequency $f$ and the energy of the upper level $E_u$ are listed in the fifth and  sixth columns, the measured velocity $V_{\rm LSR}$ 
along with its error and the 
line width $FWHM$ are shown in the seventh and eighth columns. 
The last ninth column lists the sensitivity coefficient $Q$~-- for methanol  from \cite{J11}, while for acetaldehyde $Q = 1$.
We also included the laboratory frequency uncertainties 
in the velocity uncertainty budget,
since it is the main source of errors in the $\Delta\mu/\mu$ calculations.
These type of uncertainties were not considered in \cite{Bhat} because they were unnecessary.

Since the acetaldehyde transitions have the same sensitivity coefficient, $Q=1$, their velocities can be averaged: Table~\ref{T7} shows the weighted mean velocity 
$\langle V_{\rm LSR} \rangle$ for acetaldehyde CH$_3$CHO with the corresponding error, as well as the weighted mean line width $\langle FWHM\rangle$ for comparison with the widths of methanol lines.

%---------------------Table 7
\begin{table*}[h!]
\centering
\caption{Selected CH$_3$OH and CH$_3$CHO transitions from \cite{Bhat, data}. 
Given in parenthesis are errors in the last digits.  Laboratory frequencies for CH$_3$OH and CH$_3$CHO molecules are taken from Refs. \cite{Mull} and \cite{Kle96}, respectively.}
\label{T7}
\begin{tabular}{l l c r@.l c r@.l c r@.l}
\hline
\\[-8pt]
\multicolumn{1}{c}{Object} &
\multicolumn{1}{c}{Molecule} &
Transition &
\multicolumn{2}{c}{$f$,} &
$E_u$, &
\multicolumn{2}{c}{$V_{\rm LSR}$,} &
$FWHM$, &
\multicolumn{2}{c}{$Q$} \\
&&$J_{K_u} \to J_{K_{\ell}}$ &
\multicolumn{2}{c}{MHz} & K &
\multicolumn{2}{c}{km~s$^{-1}$} &
\multicolumn{2}{c}{km~s$^{-1}$} & \\
\hline\\[-8pt]
L1544&CH$_3$OH&$5_{-1}\to4_0 E$&84521&169(10)&40.4&7&24(4)&0.44&
$-3$&6\\[2pt]
&CH$_3$CHO&$5_1\to4_1A^+$&93580&859(100)&15.8&7&14(30)&0.44&1&0\\
&CH$_3$CHO&$5_{-1}\to4_{-1}E$&93595&276(100)&15.8&7&19(30)&0.34&1&0\\
&CH$_3$CHO&$5_0\to4_0A^+$&95963&380(180)&13.8&7&18(60)&0.44&1&0\\
&CH$_3$CHO&$5_1\to4_1E$&98863&328(40)&16.6&7&11(12)&0.53&1&0\\
&CH$_3$CHO&$5_1\to4_1A^-$&98900&948(40)&16.5&7&17(12)&0.34&1&0\\
\multicolumn{6}{r}{$weighted\,\, mean\,\, \langle V_{\rm LSR} \rangle$:} & 7&14(15)\\
\multicolumn{8}{r}{$weighted\,\, mean\,\, \langle FWHM \rangle$:} & 0.42\\[2pt]

Barnard-1&CH$_3$OH&$0_{0}\to1_{-1} E$&108893&960(7)&13.1&6&565(20)&1.37&4&6\\[2pt]
&CH$_3$CHO&$5_1\to4_1A^+$&93580 &859(100)&15.8&6&57(30)&1.43&1&0\\
&CH$_3$CHO&$5_{-1}\to4_{-1}E$&93595&276(100)&15.8&6&58(30)&1.57&1&0\\
&CH$_3$CHO&$5_0\to4_0A^+$&95963&380(180)&13.8&6&66(60)&1.46&1&0\\
&CH$_3$CHO&$5_{-2}\to4_{-2}E$&96425&618(40)&22.9&6&63(16)&1.66&1&0\\
&CH$_3$CHO&$5_{2}\to4_{2}E$&96475&536(40)&23.0&6&52(15)&0.97&1&0\\
&CH$_3$CHO&$5_1\to4_1E$&98863&328(40)&16.6&6&51(12)&1.04&1&0\\
&CH$_3$CHO&$5_1\to4_1A^-$&98900&948(40)&16.5&6&55(13)&1.25&1&0\\
\multicolumn{6}{r}{$ weighted\,\, mean\,\, \langle V_{\rm LSR} \rangle$:}  & 6&55(11)\\
\multicolumn{8}{r}{$ weighted\,\, mean\,\, \langle FWHM \rangle$:} & 1.34\\[2pt]

IRAS4A&CH$_3$OH&$0_{0}\to1_{-1} E$&108893&960(7)&13.1&7&402(20)&2.45&
4&6\\[2pt]
&CH$_3$CHO&$5_1\to4_1A^+$&93580&859(100)&15.8&7&32(30)&2.30&1&0\\
&CH$_3$CHO&$5_{-1}\to4_{-1}E$&93595&276(100)&15.8&7&44(30)&2.34&1&0\\
&CH$_3$CHO&$5_{-2}\to4_{-2}E$&96425&618(40)&22.9&7&54(16)&2.41&1&0\\
&CH$_3$CHO&$5_{2}\to4_{2}E$&96475&536(40)&23.0&7&17(15)&2.36&1&0\\
&CH$_3$CHO&$5_1\to4_1E$&98863&328(40)&16.6&7&36(13)&2.24&1&0\\
&CH$_3$CHO&$6_1\to5_1A^+$&112248&728(40)&21.1&7&29(12)&2.56&1&0\\
&CH$_3$CHO&$6_{-1}\to5_{-1}E$&112254&520(40)&21.2&7&27(12)&3.06&1&0\\
\multicolumn{6}{r}{$ weighted\,\, mean\,\, \langle V_{\rm LSR} \rangle$:} & 7&32(8) \\
\multicolumn{8}{r}{$ weighted\,\, mean\,\, \langle FWHM \rangle$:} & 2.47 \\
\hline\\[-5pt]
%\multicolumn{9}{l}{\footnotesize $^a$ Laboratory frequencies CH$_3$OH are taken from \cite{Mull}.}\\
%\multicolumn{9}{l}{\footnotesize $^b$ Laboratory frequencies CH$_3$CHO are taken from \cite{Kle96}.}\\
\end{tabular}
\end{table*}

Table~\ref{T7} demonstrates 
that the widths of the methanol lines and the mean values of the  widths of 
the acetaldehyde lines are almost identical, as well as the energies of the upper levels
which differ by no more than 30~K in L1544 and no more than 10~K 
 in Barnard-1 and IRAS4A.

The relative intensities of the methanol and acetaldehyde lines were also
 utilized in \cite{Bhat} 
to calculate the excitation temperatures ($T_{\rm ex}$)
characterizing the distribution of molecules over energy levels.
Moreover, for the co-spatially distributed molecules, 
$T_{\rm ex}$ should be approximately the same.
The estimates of $T_{\rm ex}$ were carried out in \cite{Bhat} by two methods~--
Rotational Diagrams and Monte Carlo Markov Chain fitting.
In both cases, close values of the excitation temperatures were found, which may also indicate the
origin of the emission in the CH$_3$OH and CH$_3$CHO lines from the same regions.

Using the measured radial velocities
$V_{\rm LSR}$ (CH$_3$OH)  and $\langle V_{\rm LSR} \rangle$ (CH$_3$CHO)
and the corresponding $Q$-values, the following 
results were obtained:  $\Delta\mu/\mu = (0.7\pm 1.1) \times 10^{-7}$ (L1544), $\Delta\mu/\mu = (-0.1\pm 1.0) \times 10^{-7}$ 
(Barnard-1), and  $\Delta\mu/\mu = (-0.8\pm 0.8) \times 10^{-7}$
(IRAS4A).
Thus, no signal was detected at the level of $(0.8-1.1)\times10^{-7}$ in these molecular clouds. 
The slightly higher values of the upper limits on
$\Delta\mu/\mu$
for clouds L1544 and Barnard-1 are due to 
 a larger uncertainty (180 kHz) of the laboratory frequency
of the $5_0 \to 4_0 A^+$ line of acetaldehyde.
If we remove this line from the dataset, then the following values of  $\Delta\mu/\mu$
can be obtained:  $\Delta\mu/\mu = (0.7\pm0.5)\times10^{-7}$ (L1544), and
 $\Delta\mu/\mu = (-0.1\pm0.4)\times10^{-7}$ (Barnard-1).
 Now both upper limits are at the level of a few$\times10^{-8}$.
 
Since the measurements in these three molecular clouds are independent, we can average the obtained estimates of $\Delta\mu/\mu$.
The results for the initial sample and for the reduced  sample without 
line $5_0 \to 4_0 A^+$  are as follows:
$\langle \Delta\mu/\mu \rangle = (-0.2\pm0.5)\times10^{-7}$ for the former and
$\langle \Delta\mu/\mu \rangle = (0.1\pm0.4)\times10^{-7}$ for the later case.
For the final result we take the second value. 
The constraint obtained is consistent with the LPI and gives an upper limit on its violation 
in the Galactic disk at
the level of $4\times10^{-8}$.

Thus, the analysis performed shows that CH$_3$CHO is a suitable molecule 
for differential assessments of $\mu$ in combination with CH$_3$OH.
Along with other methods, acetaldehyde allows us to achieve the strongest 
limits to date on the spatial variations of $\mu$ in the Galactic disk at large galactocentric distances.

%-----------------------------------Sect 4
\section{Conclusions}
\label{Sec4}

In this study, we calculated the sensitivity coefficients $Q$ to small changes in $\mu = m_{\rm e}/m_{\rm p}$~-- the electron-to-proton mass ratio~-- for various
torsion-rotation  transitions of CH$_3$CHO in the ground torsion state $v_t = 0$, 
which fall in the microwave range of $\Delta f = 1-50$~GHz.
The main results obtained are as follows.

\begin{enumerate}
\item
The sensitivity coefficients for the 
torsion-rotation transitions with $\Delta K = \pm 1$ vary in a narrow range $0.62 \leq Q \leq 1.4$ and all of them have positive sign. 
\item
The $1_1 \to 1_{-1} E\, (\Delta K = 2)$ transition observed in astrophysics at a frequency of 1.849 GHz has the highest sensitivity coefficient $Q = 3.61$ of all for CH$_3$CHO, that confirms the previously obtained result in \cite{J11}.
\item
Analysis of astrophysical observations shows that acetaldehyde is often
co-spatially distributed with methanol,
which allows the use of acetaldehyde and methanol lines,  having similar profiles
and excitation temperatures, 
to probe  $\Delta\mu/\mu$.
\item
An example of
a combination of the CH$_3$CHO and CH$_3$OH lines for estimating $\Delta\mu/\mu$
 is demonstrated using published data \cite{Bhat} on three molecular clouds 
located  at galactocentric distances $D_{\rm GC} \sim 8$ kpc. 
The derived most stringent upper limit on $\mu$-variation $4\times 10^{-8}$ is in line with 
previously obtained constraints  based on other molecules and  methods.
Thus, no violations of the LPI are currently seen at this level in the Galactic disk at large
galactocentric distances.  
\end{enumerate}

%-----------------------------------------
\section*{Acknowledgements}
{We thank our anonymous referee for suggestions on the manuscript which improved the presentation of the results in this paper.}

%=====================================
 %References, variant A: external bibliography
%=====================================
%


\section*{References}

\begin{thebibliography}{99}

\bibitem{Kle19} I. Kleiner,  Spectroscopy of interstellar internal rotors: an important 
tool for investigating interstellar chemistry, ACS Earth Space Chem. 
{\bf 3}, 1812 (2019).

\bibitem{Hollis} J. M . Hollis, F. J. Lovas, P. R. Jewell \& L. H. Coudert,
Interstellar antifreeze: ethylene glycol,
Astrophys. J. {\bf 571}, L59 (2002).

\bibitem{Req} M. A. Requena-Torres, J. Mart\'in-Pintado, S. Mart\'in \& M. R. Morris,
The Galactic center, the largest oxigen bearing organic molecule
repository,
Astrophys. J. {\bf 672}, 352 (2008).

\bibitem{Uz25} J.-P. Uzan, Fundamental constants: from measurement to the universe, a window
on gravitation and cosmology,
Living Rev. Relativity {\bf 26}, 6 (2025).

\bibitem{VL} S. A. Levshakov \& D. A. Varshalovich, Molecular hydrogen in the z=2.811 absorbing material toward the quasar PKS 0528-250, Mon. Not. R. Astron. Soc. 
{\bf 212}, 517 (1985).

\bibitem{KL2013} M. G. Kozlov \& S. A. Levshakov, Microwave and submillimeter molecular transitions and their dependence on fundamental constants, Ann. Phys. 
{\bf 525}, 452 (2013).

\bibitem{FK} V. V. Flambaum \& M. G. Kozlov, Limit on the cosmological variation of $m_p/m_e$ from the inversion spectrum of ammonia,
 Phys. Rev. Lett. {\bf 98}, 240801 (2007).

\bibitem{J13} P. Jansen, L.-H. Xu, I. Kleiner, H. L. Bethlem \&
W. Ubachs, Methyl mercaptan (CH$_3$SH) as a probe for
variation of the proton-to-electron mass ratio, Phys. Rev.
A {\bf 87}, 052509 (2013).

\bibitem{Il} V. V. Ilyushin, P. Jansen, M. G. Kozlov, S. A. Levshakov,
I. Kleiner, W. Ubachs \&  H. L. Bethlem, Sensitivity to
a possible variation of the proton-to-electron mass ratio
of torsion-wagging-rotation transitions in methylamine
CH$_3$NH$_2$, Phys. Rev. A {\bf 85}, 032505 (2012).

\bibitem{J11} P. Jansen, I. Kleiner, L.-H. Xu, W. Ubachs \& H. L.
Bethlem, Sensitivity of transitions in internal rotor
molecules to a possible variation of the proton-to-electron mass ratio, Phys. Rev. A
 {\bf 84}, 062505 (2011).

\bibitem{LKR} S. A. Levshakov, M. G. Kozlov \& D. Reimers,
Methanol as a tracer of fundamental constants, Astrophys. J. {\bf 738}, 26 (2011).

\bibitem{J112} P. Jansen, L.-H. Xu, I. Kleiner, W. Ubachs \& H. L.
Bethlem, Methanol as a sensitive probe for spatial and
temporal variations of the proton-to-electron mass ratio,
Phys. Rev. Lett. {\bf 106}, 100801 (2011).

\bibitem{VKL} J. S. Vorotyntseva, M. G. Kozlov \& S. A. Levshakov,
Methanol isotopologues as a probe for spatial and temporal variations of 
the electron-to-proton mass ratio,
Mon. Not. R. Astron. Soc. {\bf 527}, 2750 (2024).

\bibitem{VLK} J. S. Vorotyntseva, S. A. Levshakov \& M. G. Kozlov,  Spectroscopic shifts in deuterated methanol induced by variation of $m_e/m_p$, 
Phys. Rev. A {\bf 110}, 012802 (2024).

\bibitem{L2013} S. A. Levshakov, D. Reimers, C. Henkel, B. Winkel, A. Mignano, M. Centuri\'on \& P. Molaro, Limits on the spatial variations of the electron-to-proton mass ratio in the Galactic plane, Astron. Astrophys. {\bf 559}, A91 (2013).

\bibitem{Dap}  M. Dapr\`a, C. Henkel, S. A. Levshakov, K. M. Menten,
S. Muller, H. L. Bethlem, S. Leurini, A. V. Lapinov \&
W. Ubachs, Testing the variability of the proton-to-
electron mass ratio from observations of methanol in the
dark cloud core L1498, Mon. Not. R. Astron. Soc. {\bf 472}, 4434 (2017).

\bibitem{L22} S. A. Levshakov, I. I.  Agafonova,  C. Henkel, Kee-Tae Kim, 
M. G. Kozlov, B. Lankhaar \& W. Yang, 
Probing the electron-to-proton mass ratio gradient in the
Milky Way with Class~I methanol masers, Mon. Not. R. Astron. Soc.
{\bf  511}, 413 (2022).

\bibitem{Ell} S. Ellingsen, M. Voronkov \& S. Breen, Practical limitations on astrophysical observations of methanol to investigate variations in the proton-to-electron mass ratio, Phys. Rev. Lett. {\bf 107}, 270801 (2011).

\bibitem{VL2024} J. S. Vorotyntseva \& S. A. Levshakov, Torsion-rotational 
transitions in methanol as a probe
 of fundamental physical constants~-- electron and proton masses, 
 JETP Lett. {\bf 119}, 649 (2024).

\bibitem{VL2025} J. S. Vorotyntseva \& S. A. Levshakov, Indication of the electron-to-proton mass ratio variation in the Galaxy, JETP Lett. {\bf 121} 589 (2025).

\bibitem{VLH2025} J. S. Vorotyntseva,  S. A. Levshakov \& C. Henkel, Indication of the electron-to-proton mass ratio variation in the Galaxy. II. 3~mm
methanol lines towar Sgr B2(N) and
B2(M) molecular clouds,
JETP Lett. {\bf 122} in press (2025).

\bibitem{RF} D. Rabli \& D. R. Flower, The rotational structure of
methanol and its excitation by helium, Mon. Not.
R. Astron. Soc. {\bf 403}, 2033 (2010).

\bibitem{NB} H. Nadgaran \& J. G. Baker, Comparison between single-state and global fits to the millimeter wave spectra of acetaldehyde in excited torsional states, ASP Conf. Series {\bf 81}, 332 (1995).

\bibitem{TowSh} C. Townes \& A. Schawlow, Microwave spectroscopy
 (McGraw-Hill, New York, 1955).

\bibitem{Kle96} I. Kleiner, F .J. Lovas \& M. Godefroid, Microwave spectra of molecules of astrophysical Interest. XXIII. 
Acetaldehyde, J. Phys. Chem. Ref. Data {\bf 25}, No.4 (1996).

\bibitem{Bell} M. B. Bell, H. E. Matthews \&  P. A. Feldman,
Observations of microwave transitions of A-state acetaldehyde in SGR B2, 
Astron. Astrophys. {\bf 127}, 420 (1983).

\bibitem{Hol} J. M. Hollis, A. Remijan, P. R. Jewell \& F. J. Lovas, 
Cyclopropenone (c-H$_2$C$_3$O): a new interstellar ring molecule,
Astrophys. J. {\bf 642}, 933 (2006).

\bibitem{Fou} N. Fourikis, 
M. W. Sinclair,  B. J. Robinson, P. D. Godfrey \& R. D. Brown, 
Microwave emission of the $2_{11} \to  2_{12}$ rotational transition in interstellar acetaldehyde,
Aust. J. Phys. {\bf 27}, 425 (1974).

\bibitem{Matt85} H. E. Matthews, P. Friberg \& W. M. Irvine, 
The detection of acetaldehyde in cold dust clouds,
Astrophys. J. {\bf 290}, 609 (1985).

\bibitem{Sci24} S. Scibelli,  Y.  Shirley, A. Meg\'ias \& I. Jim\'enez-Serra,
Survey of complex organic molecules in starless and pre-stellar cores in the Perseus molecular cloud, Mon. Not. R. Astron. Soc. {\bf 533}, 4104 (2024).

\bibitem{Kaif} N. Kaifu, 
M. Ohishi, K. Kawaguchi, S. Saito, S. Yamamoto, T. Miyaji, K. Miyazawa, 
S.-I. Ishikawa, C. Noumaru, S. Harasawa, M. Okuda \& H. Suzuki, 
A 8.8--50GHz complete spectral line survey toward TMC-1 I. Survey data,
Pub. Astron. Soc. Japan {\bf 56}, 69 (2004).

\bibitem{Soma} T. Soma, N. Sakai, Y Watanabe \& S. Yamamoto, Complex organic molecules in Taurus Molecular Cloud-1, Astrophys. J. {\bf 854}, 116 (2018).

\bibitem{Sai} S. Saito, 
S. Yamamoto, K. Kawaguchi, M. Ohishi,  H. Suzuki, S.-I. Ishikawa, N. Kaifu, 
The microwave spectrum of the CP radical and related astronomical search,
Astrophys. J. {\bf 341}, 1114 (1989).

\bibitem{Bac} A. Bacmann, V. Taquet, A. Faure,  C. Kahane \&  C. Ceccarelli,
Detection of complex organic molecules in a prestellar core: a new challenge for astrochemical models, Astron. Astrophys. {\bf 541}, L12 (2012).

\bibitem{Bus} L. A. Busch, A. Belloche, R. T. Garrot, H. S. P. M\"uller \& K. M. Menten,
Shocking Sgr B2 (N1) with its own outflow. A new perspective on segregation between O- and N-bearing molecules,
Astron. Astrophys. {\bf 681}, A104 (2024).

\bibitem{Sci20} S. Scibelli \& Y. Shirley, 
Prevalence of complex organic molecules in starless and prestellar cores within the Taurus Molecular Cloud, Astrophys. J. {\bf 891}, 73 (2020).

\bibitem{Cod} C. Codella,  C. Ceccarelli, E. Bianchi,  N. Balucani, L. Podio,  
P. Caselli,  S. Feng,  B. Lefloch,  A. L\'opez-Sepulcre,  R. Neri,  
S. Spezzano \& M. De Simone, 
Seeds of life in space (SOLIS). V. Methanol and acetaldehyde in the protostellar jet-driven shocks L1157-B0 and B1,
Astron. Astrophys. {\bf 635}, A17 (2020).

\bibitem{Cer} J. Cernicharo,  N. Marcelino, E. Roueff,  M. Gerin, 
A. Jim\'enez-Escobar, G. M. Mu\~noz Caro,
Discovery of the methoxy radical, CH$_3$O, toward B1: dust grain and gas-phase chemistry in cold dark clouds, 
Astrophys. J. Lett. {\bf 759}, L43 (2012).

\bibitem{Bhat} B. Bhat, R. Kar, S. K. Mondal, R. Ghosh,  P. Gorai, Prasanta, T. Shimonishi, Kei E. I. Tanaka, K. Furuya \& A. Das,
Chemical evolution of some selected complex organic molecules in low-mass star-forming Regions, Astrophys. J. {\bf 958}, 111 (2023).

\bibitem{Char} S. B. Charnley, 
Acetaldehyde in star-forming regions,
Adv. Space Research {\bf  33}, 23 (2004).

\bibitem{Sci21} S. Scibelli, Y. Shirley,  A. Vasyunin \& R. Launhardt, 
Detection of complex organic molecules in young starless core L1521E,
Mon. Not. R. Astron. Soc.  {\bf 504}, 5754 (2021).

\bibitem{Nagy} Z. Nagy,  S. Spezzano, P. Caselli,  A. Vasyunin,  
M. Tafalla,  L. Bizzocchi,  D. Prudenzano \&   E. Redaelli,
 The chemical structure of the very young starless core L1521E,
Astron. Astrophys. {\bf 630}, A136 (2019).

\bibitem{Vas} C. Vastel, C. Ceccarelli,  B. Lefloch \&  R. Bachiller, 
The origin of complex organic molecules in prestellar cores,
Astrophys. J. Lett. {\bf 795}, L2 (2014).

\bibitem{Berg} J. B. Bergner, K. I. \"Oberg,  R. T. Garrod \& D. M. Graninger, 
Complex organic molecules toward embedded low-mass protostars,
Astrophys. J. {\bf 841}, 120 (2017).

\bibitem{Obe} K. I. \"Oberg, T. Lauck \& D. Graninger, 
Complex organic molecules during low-mass star formation: pilot survey results,
Astrophys. J. {\bf 788}, 68 (2014).

\bibitem{Hsi} T.-H. Hsieh, J. E. Pineda,  D. M. Segura-Cox,  
P. Caselli, M. T. Valdivia-Mena, C. Gieser, M. J. Maureira, A. Lopez-Sepulcre, 
L.  Bouscasse, R. Neri, Th. M\"oller, A. Dutrey, A. Fuente,  
D. Semenov, E. Chapillon, N. Cunningham, Th. Henning, V. Pi\'etu, 
I. Jimenez-Serra,  S. Marino \& Ceccarelli, 
PRODIGE - envelope to disk with NOEMA. III. The origin of complex organic molecule emission in SVS13A, Astron. Astrophys. {\bf 686}, A289 (2024).

\bibitem{Fay} E. C. Fayolle,
K. I. \"Oberg,  R. T. Garrod, E. F. van Dishoeck \& S. E. Bisschop, 
Complex organic molecules in organic-poor massive young stellar objects,
Astron. Astrophys. {\bf 576}, A45 (2015).

\bibitem{Lyk} J. M. Lykke, A. Coutens, 
J. K. J\o rgensen,  
M. H. D. van der Wiel,  R. T. Garrod, H. S. P. M\"uller, P. Bjerkeli,  T. L. Bourke,  
H. Calcutt, M. N. Drozdovskaya, C. Favre,  E. C. Fayolle,  
S. K. Jacobsen,  K. I. \"Oberg, M. V. Persson, E. F. van Dishoeck \& S. F. Wampfler, 
The ALMA-PILS survey: first detections of ethylene oxide, acetone and propanal toward the low-mass protostar IRAS 16293-2422,
Astron. Astrophys. {\bf 597}, A53 (2017).

\bibitem{Mull11}  
S. Muller,  A. Beelen, M. Gu\'elin,  S. Aalto, J. H. Black,  F. Combes,  
S. J. Curran, P. Theule \& S. N. Longmore, 
Molecules at z = 0.89. A 4-mm-rest-frame absorption-line survey toward PKS 1830-211,
Astron. Astrophys. {\bf 535}, A103 (2011). 

\bibitem{data} The data that support the findings of this article are openly available at https://doi.org/10.3847/1538-4357/acfc4d.

\bibitem{Mull} H. S. P. M\"uller, K. M.  Menten \&  H. M\"ader,
Accurate rest frequencies of methanol maser and dark cloud lines, 
Astron. Astrophys {\bf 428}, 1019 (2004).


\end{thebibliography}
\end{document}